\documentclass[a4paper]{article}
\usepackage{INTERSPEECH2022}
\usepackage[colorlinks,linkcolor=blue]{hyperref} 
\usepackage{float}
\usepackage{placeins}
\usepackage{multirow}

\title{Backend Ensemble for Speaker Verification and Spoofing Countermeasure}
\name{Li Zhang, Yue Li, Huan Zhao, Qing Wang, Lei Xie$^{*}$ \thanks{* Corresponding author.}}
\address{
   Audio, Speech and Language Processing Group (ASLP@NPU), School of Computer Science, Northwestern Polytechnical University (NPU), Xi'an, China}
\email{lizhang.aslp.npu@gmail.com, lxie@nwpu.edu.cn}

\begin{document}

\maketitle
\begin{abstract}
This paper describes the NPU system submitted to Spoofing Aware Speaker Verification Challenge 2022. We particularly focus on the \textit{backend ensemble} for speaker verification and spoofing countermeasure from three aspects. Firstly, besides simple concatenation, we propose circulant matrix transformation and stacking for speaker embeddings and countermeasure embeddings. 
With the stacking operation of newly-defined circulant embeddings, we almost explore all the possible interactions between speaker embeddings and countermeasure embeddings. Secondly, we attempt different convolution neural networks to selectively fuse the embeddings’ salient regions into channels with convolution kernels. Finally, we design parallel attention in 1D convolution neural networks to learn the global correlation in channel dimensions as well as to learn the important parts in feature dimensions. Meanwhile, we embed squeeze-and-excitation attention in 2D convolutional neural networks to learn the global dependence among speaker embeddings and countermeasure embeddings. Experimental results demonstrate that all the above methods are effective. After fusion of four well-trained models enhanced by the mentioned methods, the best SASV-EER, SPF-EER and SV-EER we achieve are 0.559\%, 0.354\% and 0.857\% on the evaluation set respectively. Together with the above contributions, our submission system achieves the fifth place in this challenge. 
\end{abstract}

\section{Introduction}
The key of automatic speaker verification~(ASV) is to decide whether 
two speech utterances come from the same speaker~\cite{bimbot2004tutorial,naika2018overview}. With the development of deep neural network (DNN) and easy availability of computing resources and massive data, ASV technology has delivered the high accuracy required in voice-enabled IoT gadgets control, speech authorization and forensic applications~\cite{desplanques2020ecapa,giraldo201918muw,machado2019forensic}. However, ASV systems are vulnerable under various kinds of malicious spoofing attacks, i.e., specially crafted utterances generated by adversaries to deceive the ASV system and to provoke false accepts~\cite{jung2022sasv,wang2020asvspoof,mittal2021automatic,wang2020inaudible}. 

The data scenarios for speaker spoofing include logical access~(LA), physical access~(PA), speech deep fake ~(DF)~\cite{yamagishi2021asvspoof}.
The generated human-like speech deceiving ASV systems poses a great threat to the security of society if misused malignantly. Fortunately, this problem has 
attracted the attention of many researchers. There are several anti-spoofing challenges~\cite{delgado2021asvspoof,yi2022add} held to boost the development of countermeasure~(CM) systems to help detecting spoofing attacks~\cite{tak2021end,hanilcci2018linear, chadha2021review,li2019anti}. While most CM systems achieve fabulous performance in detecting spoofing speech, they seriously affect the performance of the zero-effort impostors' detection when they work with ASV systems~\cite{kinnunen2020tandem}. Spoofing Aware Speaker Verification Challenge 2022~(SASV)~\cite{jung2022sasv} provides a common platform for researchers to extend the focus of ASVspoof upon CMs to the consideration of integrated systems 
where both CM and ASV subsystems are optimized jointly to improve reliability~\cite{jung2022sasv}. The SASV challenge focuses on spoofing attacks generated using speech synthesis/text-to-speech (TTS) and voice conversion (VC), which are LA spoofing. It proposes to jointly optimize the automatic ASV system and spoofing CM system, which aims to develop a new spoofing-aware speaker verification system.

Recently, several works pay attention to the ensemble of ASV and CM systems considering both performances of the two systems. The ensemble methods can be classified into three types, which are the tandem for ASV and CM systems to make a decision of speaker verification, multi-task learning frameworks for ASV and CM systems, backend ensemble for ASV and CM systems. In terms of the tandem for ASV and CM systems, Tomi et al.~\cite{kinnunen2020tandem} proposed several new tandem detection cost function (t-DCF) to assess the reliability of spoofing CMs deployed in tandem with an ASV system. Kanervisto et al.~\cite{kanervisto2021optimizing} optimized the tandem system directly by creating a differentiable version of t-DCF and employing techniques from reinforcement learning. For the multi-task learning frameworks for ASV and CM systems, Li et al.~\cite{li2019multi} proposed a multi-task learning neural network to make a joint system of ASV and anti-spoofing, which verified that joint optimization was more advantageous than the cascaded systems. Moreover, Li et al.~\cite{li2020joint} proposed a multi-task learning framework with contrastive loss to joint decision of anti-spoofing and ASV. As for the backend ensemble for ASV and CM systems, Gomez-Alanis et al.~\cite{gomez2020joint} developed an integration neural network and a loss function based on the minimization of the area under the expected (AUE) performance and spoofability curve~(EPSC) to jointly process the embeddings extracted from ASV and anti-spoofing systems. Chettri et al.~\cite{chettri2019ensemble} combined deep neural networks and traditional machine learning models as ensemble models by logistic regression to integrate spoofing detection in an ASV system.

In this challenge, we focus on exploring the backend embedding ensemble of ASV and CM systems.
Our target is to build a backend ensemble system trained with speaker embeddings and CM embeddings to make a joint decision of speaker verification. The primary work of the backend ensemble system is to mine as much effective speaker discrimination information and spoofing detection information as possible from speaker embeddings and CM embeddings.
 Specifically, we exploit different embedding fusion methods, neural network frameworks and enhanced attention mechanisms to aggregate valuable information for spoofing aware speaker verification. Together with the proposed methods, our experimental results on the evaluation set have significant improvements compared with the baseline results in the challenge. 


\section{System Overview}
\begin{figure}[th]
  \centering
  \includegraphics[width=\linewidth]{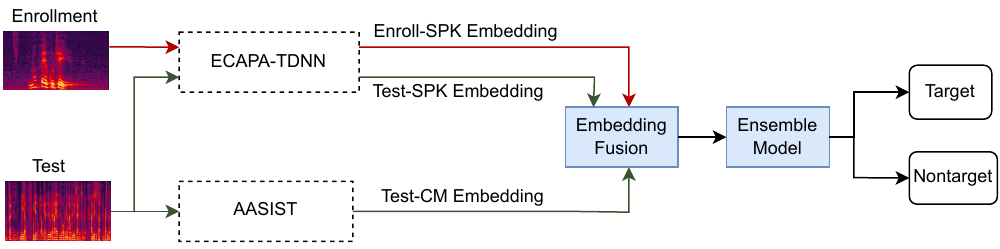}
  \caption{The overview of backend ensemble system. }
  \label{fig:system_overview}
\end{figure}

Figure~\ref{fig:system_overview} gives an overview of our system. It consists of three modules which are embedding extractors~(ECAPA-TDNN and AASIST), embedding fusion and ensemble model. The outputs are labels of test trials denoting whether the enrollment and test utterances belong to the same speaker.  Our contributions are to explore different embedding fusion methods, ensemble network frameworks and attention mechanisms in the ensemble modules to 
learn shared speaker and detection information from ASV and CM systems for spoofing aware speaker verification.

In the embedding extractor module, the speaker embedding extractor and CM embedding extractor we use are pretrained ECAPA-TDNN~\cite{desplanques2020ecapa} and AASIST~\cite{jung2021aasist} offered by the challenge organizer~\cite{jung2022sasv}. Therefore, the two embedding extractors in the dashed boxes do not participate in training in our system.

In the embedding fusion module, we exploit three kinds of embedding fusion methods which are concatenation, circulant matrix transformation and stacking.

\begin{figure}[th]
  \centering
  \includegraphics[width=\linewidth]{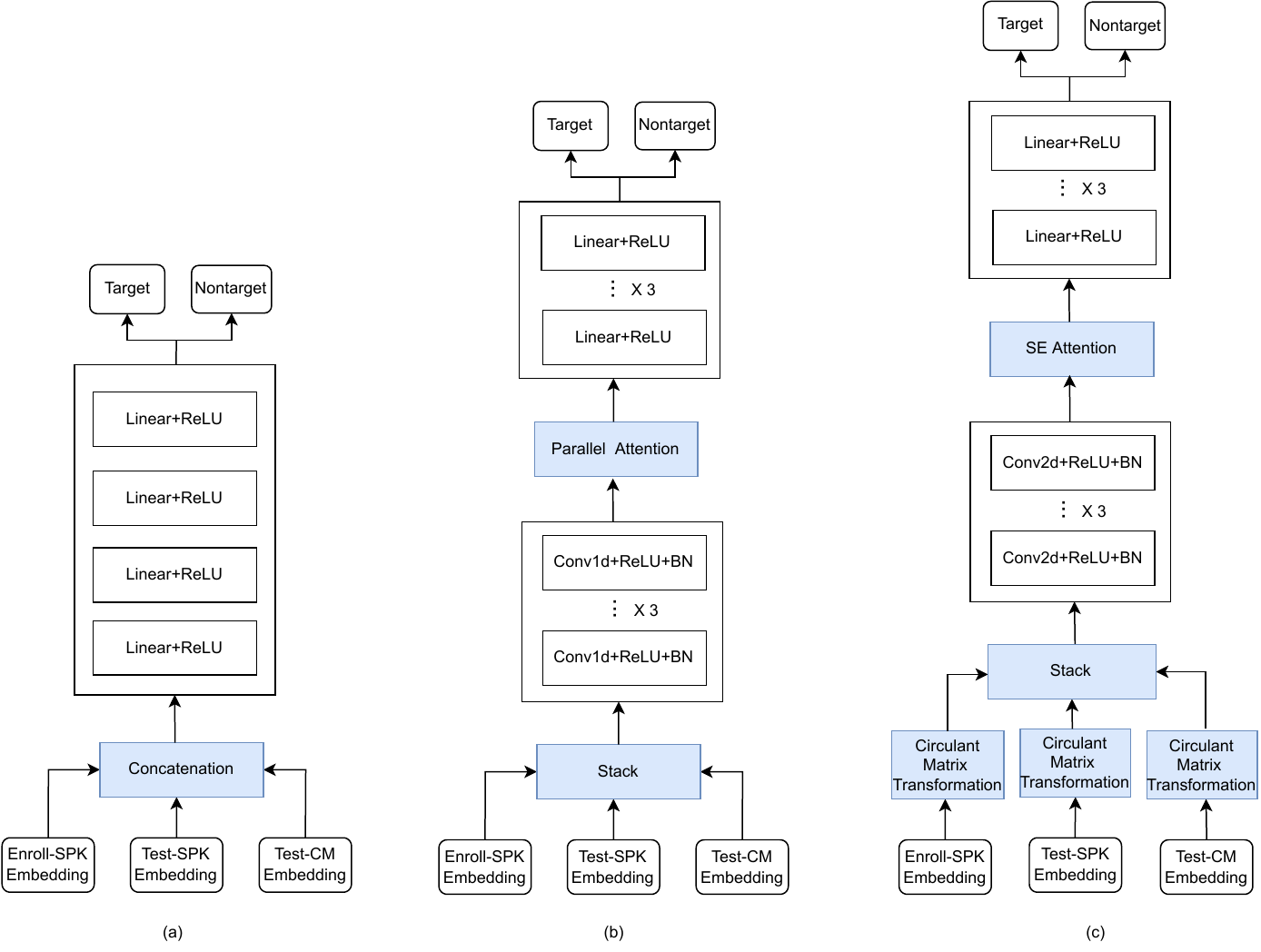}
  \caption{Different backend ensemble modules: (a) is the enlarged baseline model~(DNN), (b) is the 1D CNN with parallel attention, and (c) is the circulant matrix transformation for 2D CNN with SE attention. }
  \label{fig:different-frameworks}
\end{figure}

For the ensemble model module illustrated in Figure~\ref{fig:different-frameworks}, besides the enlarged DNN baseline, we introduce another two backend frameworks based on convolution neural networks~(CNN) . Firstly, we enlarge the baseline2 model which is a 3-layer DNN~\cite{jung2022sasv}, but the performance saturates as the model gets deeper. 
Then we stack the embeddings and use 1D CNN to derive the decision of speaker verification. Meanwhile, we propose parallel attention after convolutional blocks to learn the global relationship in channel dimensions as well as to learn the important regions in feature dimensions. Finally, we propose to make circulant matrix transformation on embeddings and stack newly-defined embeddings before feeding them into a 2D CNN with squeeze-and-excitation~(SE) attention~\cite{hu2018squeeze,hou2021coordinate}. 

\section{Backend Ensemble Frameworks}
The performance of the baseline2 model provided by challenge organizer~\cite{jung2022sasv} is easy to saturate as the model gets deeper. 
Therefore, we propose another two CNN-based backend ensemble frameworks for speaker verification and spoofing countermeasure. 
 
\subsection{1D CNN with Parallel Attention}
\begin{figure}[th]
  \centering
  \includegraphics[width=\linewidth]{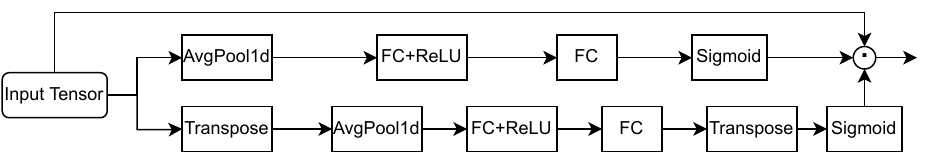}
  \caption{The parallel attention~(PA) in 1D CNN.}
  \label{fig:global-attention}
\end{figure}

We adopt 1D CNN to process the stacked speaker embeddings and CM embeddings. The aim is to use convolutional kernels for selectively fusing the different embedding regions into channels. Then we introduce a parallel attention~(PA) module to learn global interactions among different embeddings and focus on important areas in the feature dimension, as illustrated in Figure~\ref{fig:global-attention}. The PA module learns attention masks along with the channel dimension and feature dimension respectively.

Suppose the input tensor generated from 1D convolutional blocks is $T^{(B \times C \times F)}$, where $B$, $C$, $F$ are batchsize, channel dimension and feature dimension respectively. We split two paths to calculate the channel-wise and feature-wise attention at the same time. The average pooling on the feature dimension and channel dimension of parallel attention is calculated as:

\begin{scriptsize}
\begin{eqnarray}
    T^{B \times F }=\frac{1}{C}\sum_{i=1}^{C} T^{(B \times C \times F)},
\label{SE1_Avg1}
\end{eqnarray}
\end{scriptsize}

\begin{scriptsize}
\begin{eqnarray}
    T^{B \times C }=\frac{1}{F}\sum_{i=1}^{F} \tau {(T^{(B \times C \times F)})},
\label{SE1_Avg2}
\end{eqnarray}
\end{scriptsize}

\noindent where $\tau$ in Eq.~\ref{SE1_Avg2} is to transpose the channel dimension and feature dimension in $T^{B \times C \times F }$. Then we use linear connection layers to learn the channel-wise and feature-wise attention in parallel. The equations are:

\begin{scriptsize}
\begin{eqnarray}
    T_1^{B \times F}=\rho( ( T^{B \times F} \otimes  W_1^{F \times F/r}) \otimes  W_2^{F/r \times F} ),
\label{ESE2}
\end{eqnarray}
\end{scriptsize}
\begin{scriptsize}
\begin{eqnarray}
    T_2^{B \times C}=\rho( (T^{B \times C} \otimes W_3^{C \times C/r} )\otimes W_4^{C/r \times C} ),
\label{ESE1}
\end{eqnarray}
\end{scriptsize}

\noindent where $\otimes$ is matrix multiplication, $r$ is the middle layer dimension reduction factor and $\rho$ is the $sigmoid$ operation. The weighted intermediate tensor $T^{B \times C \times F }$ is calculated as:

\begin{scriptsize}
 \begin{eqnarray}
    (T^{B \times C \times F })^{'}= T_2^{B \times C} \cdot T^{B \times C \times F} \cdot   T_1^{B \times F},
\label{SE1_Final}
\end{eqnarray}
\end{scriptsize}

\noindent where $\cdot$ is dot products with automatic dimension expansion. $(T^{B \times C \times F })^{'}$ is the output tensor of the PA module.

\subsection{Circulant Matrix Transformation for 2D CNN with SE Attention}
 To further use convolutional kernels for selectively fusing the embeddings' salient regions, we adopt circulant matrix transformation to transfer speaker embeddings and CM embeddings into circulant matrices. The circulant matrix is a square matrix, in which all row vectors are composed of the same elements and each row vector is rotated one element to the right relative to the preceding row vector~\cite{gray2006toeplitz}. Suppose the enrollment speaker embedding, test speaker embedding and test CM embedding are $e_{i} = \{x_{1},x_{2},x_{3}...x_{d}\}_{1:d}$, $t_{i} = \{s_{1},s_{2},s_{3}...s_{b}\}_{1:b}$ and $c_{i} =\{m_{1},m_{2},m_{3}...m_{q}\}_{1:q}$, the circulant matrix transformation are formulated as follows.
\begin{scriptsize}
\begin{equation}
{e_i}^{'}=\left(\begin{array}{ccccc}x_{1} & x_{2} & x_{3} & \cdots & x_{d} \\ x_{d} & x_{0} & x_{1} & x_{2} & \cdots \\ x_{d-1} & x_{d} & x_{1} & \cdots & \\ \ddots & \ddots & \ddots & \ddots & \ddots \\ x_{1} & x_{2} & \cdots & x_{d} & x_{0}\end{array}\right),
\label{e1}
\end{equation}
\end{scriptsize}

\begin{scriptsize}
\begin{equation}
{t_i}^{'}=\left(\begin{array}{ccccc}s_{1} & s_{2} & s_{3} & \cdots & s_{b} \\ s_{b} & s_{0} & s_{1} & s_{2} & \cdots \\ s_{b-1} & s_{b} & s_{1} & \cdots & \\ \ddots & \ddots & \ddots & \ddots & \ddots \\ s_{1} & s_{2} & \cdots & s_{b} & s_{0}\end{array}\right),
\label{e2}
\end{equation}
\end{scriptsize}
\begin{scriptsize}
\begin{equation}
{c_i}^{'}=\left(\begin{array}{ccccc}m_{1} & m_{2} & m_{3} & \cdots & m_{q} \\ m_{q} & m_{0} & m_{1} & m_{2} & \cdots \\ m_{q-1} & m_{q} & m_{1} & \cdots & \\ \ddots & \ddots & \ddots & \ddots & \ddots \\ m_{1} & m_{2} & \cdots & m_{q} & m_{0}\end{array}\right).
\label{e3}
\end{equation}
\end{scriptsize}
As each row of the circulant matrix shifts one element, with newly-defined interaction operations, we almost explore all possible interactions among different embeddings. After that, we stack the three circulant matrices into one tensor and feed it into a 2D CNN. Meanwhile, we embed the SE attention~\cite{hu2018squeeze} after the last convolutional layer to learn the dependency of different embeddings.

\section{Experimental Setup}
\subsection{Datasets}
All the experiments presented further are conducted on the logical access~(LA) training, development and evaluation sets in ASVspoof 2019 database~\cite{todisco2019asvspoof}. The assessment results are derived from ASV trials in development and evaluation sets~\cite{jung2022sasv}. 

\subsection{Model Configurations}
In this paper, eight models are trained with the following configurations.
\begin{itemize}
    \item \textbf{Extend(512)\_DNN} The neural nodes of 4-layer DNN are [512, 256, 128, 64].
    \item 
    \textbf{Extend(1024)\_DNN} The neural nodes of 5-layer DNN are [1024, 512, 256, 128, 64].
    
    \item \textbf{1D\_CNN} There are 3-layer 1D convolution, 1-layer adaptive average pooling and 3-layer DNN in the 1D\_CNN model. The channels and kernels of 3-layer 1D convolution are [256, 128, 64] and [3, 3, 3]. Each convolutional layer is followed with a normalization layer and a LeakReLU activation layer. The neural nodes of 3-layer DNN are [512, 256, 64]. 
    
    \item \textbf{1D\_CNN\_SE} The SE attention layer is embedded after the third convolutional network layer in 1D\_CNN. The inner channel reduction ratio of SE attention is eight. Other configurations of this model are the same as those in the 1D\_CNN model.
    \item \textbf{1D\_CNN\_PA}
    The parallel attention layer is embeded after the third convolutional network layer in 1D\_CNN. The inner channel reduction ratio of parallel attention is eight. Other configurations of this model are the same as those in the 1D\_CNN model.
    \item \textbf{2D\_CNN} There are 4-layer 2D convolution, 1-layer adaptive average pooling and 3-layer DNN in the 2D\_CNN model. The channel and kernel configures are [32, 64, 128, 256] and [5, 3, 3, 3]. The configuration of adaptive average pooling is [16, 16]. The neural nodes of DNN are [256, 128, 64]. 
    
    \item \textbf{2D\_CNN\_SE} The SE attention~\cite{hu2018squeeze} layer is embedded after the third convolutional network layer. Other configurations of this model are the same as those in the 2D\_CNN model. 
    \item \textbf{2D\_CNN\_VSE}
    In addition to SE attention, we also use a variant of SE attention~\cite{hou2021coordinate} in 2D\_CNN model. The variant SE attention~(VSE) has been demonstrated to be effective in speaker verification~\cite{zhang2021duality}. Other configurations of this model are the same as those in the 2D\_CNN model.
\end{itemize}

\subsection{Training Setup}
We adopt the well pre-trained ECAPA-TDNN~\cite{desplanques2020ecapa} and AASIST~\cite{jung2021aasist} models provided by the challenge organizer~\cite{jung2022sasv} as our embedding extractors. In the training step, the embedding extractors are fixed without joint training. The initial learning rate is 1e-3. The optimizer is Adam with keras scheduler and weight decay is 1e-3. The loss function is cross entropy with bias-weight [0.1, 0.9] because of the unbalance of bonafide and spoofing datasets in ASVspoof~2019 train set.  
\subsection{Score Metric}
The classical equal error rate~(EER)~\cite{cheng2004method} is used as the primary metric. Specifically, there are three kinds of EER metrics, i.e., SASV-EER, SPF-EER and SV-EER~\cite{jung2022sasv1}. The SASV-EER does not distinguish between different speaker
(zero-effort, non-target, or impostor) access attempts and spoofed access attempts. SPF-EER is the metric of spoofing attacks and target trials. SV-EER is the metric of traditional ASV trials without spoofing attacks~\cite{jung2022sasv}.

\section{Experimental Results}
Experimental results in Table~\ref{tab:results_train} are derived from the models trained with ASVspoof~2019 training set. We can see that the SASV-EER of Extend~(512)\_DNN model has an absolute reduction of 1.529\% compared with that of baseline2 model~\cite{jung2022sasv} on the evaluation set. After we stack speaker embeddings and CM embeddings and feed them into the 1D\_CNN model, the SASV-EER on the evaluation set dramatically reduces to 1.361\%. The performance improvements are thanks to the decline in SV-EER. But the SPF-EER of 1D\_CNN on evaluation set slightly rises. When we embed SE attention into 1D\_CNN model, the SPF-EER gets more reduction. Meanwhile, the SASV-EER of the evaluation set is reduced by 0.244\% compared with that of 1D\_CNN. After we replace the SE attention with the parallel attention in 1D\_CNN\_SE, the SASV-EER on the evaluation set is further reduced by 0.093\%. With the help of circulant matrix transformation, the SASV-EER of 2D\_CNN model is better than that of 1D\_CNN model on the evaluation set. With adding SE attention in 2D\_CNN, the SASV-EER of 2D\_CNN\_SE model even reaches 0.998\% on the evaluation set, and we get the lowest SPF-EER and SV-EER compared with other models in Table~\ref{tab:results_train}. Moreover, when we embed the VSE attention in 2D\_CNN model, the SASV-EER of the evaluation set is decreased by 0.134\% compared with that of 2D\_CNN model.  
\begin{table*}[th]
\renewcommand\arraystretch{1.2}
\centering
\caption{EER~(\%) results~(trained with training set) on SASV 2022 development and evaluation partitions.}
\label{tab:results_train}
\begin{tabular}{cccccccc}
\hline
\multirow{2}{*}{Model Index} & \multirow{2}{*}{Model Name} & \multicolumn{3}{c}{DEV}     & \multicolumn{3}{c}{EVAL}                 \\ \cline{3-8} 
                             &                             & SASV-EER & SPF-EER & SV-EER & SASV-EER       & SPF-EER        & SV-EER \\ \hline
A                            & Model\_baseline2            & 4.85     & 0.13    & 12.87  & 6.37           & 0.78           & 11.48  \\
B                            & Extend~(512)\_DNN            & 3.973    & 0.193   & 9.097  & 4.841          & 0.797          & 8.429  \\
C                            & Extend~(1024)\_DNN           & 3.705    & 0.1634  & 9.652  & 4.926          & 0.710          & 8.683  \\
D                            & 1D\_CNN                     & 0.606    & 0.135   & 1.456  & \textbf{1.361} & 1.135          & 1.750  \\
E                            & 1D\_CNN\_SE                & 0.876    & 0.110   & 1.699  & \textbf{1.117} & 0.519          & 1.638  \\
F                            & 1D\_CNN\_PA                 & 1.022    & 0.103   & 1.954  & \textbf{1.024} & 0.819          & 1.378  \\
G                            & 2D\_CNN                     & 0.687    & 0.067   & 1.752  & \textbf{1.212} & \textbf{0.416} & 2.019  \\
H                            & 2D\_CNN\_SE                & 0.846    & 0.135   & 2.167  & \textbf{0.998} & \textbf{0.497} & 1.582   
\\
I                           & 2D\_CNN\_VSE                & 1.257    & 0.134   & 2.409  & \textbf{1.078} & 0.509 & 1.765  \\ \hline
\end{tabular}
\end{table*}
\begin{table*}[h]
\renewcommand\arraystretch{1.2}
\centering
\caption{EER~(\%) results~(trained with training and dev sets) on SASV 2022 development and evaluation partitions.}
\label{tab:results_train_dev}
\begin{tabular}{cccccccc}
\hline
\multirow{2}{*}{Model Index} & \multirow{2}{*}{Model Name} & \multicolumn{3}{c}{DEV}     & \multicolumn{3}{c}{EVAL}                         \\ \cline{3-8} 
                             &                             & SASV-EER & SPF-EER & SV-EER & SASV-EER       & SPF-EER        & SV-EER         \\ \hline
B\_Aug                       & Extend~(512)\_DNN            & 0.011    & 8.97e-5 & 0.017  & 3.026          & 0.837          & 5.497          \\
D\_Aug                       & 1D\_CNN                     & 0.011    & 0.067   & 1.666  & \textbf{0.837} & 0.350          & 1.303          \\
E\_Aug                       & 1D\_CNN\_SE                & 0.078    & 8.97e-5 & 0.134  & \textbf{0.812} & 0.570          & \textbf{0.981} \\
F\_Aug                       & 1D\_CNN\_PA                 & 0.067    & 0.066   & 0.022  & \textbf{0.768} & 0.465          & 1.053          \\
G\_Aug                       & 2D\_CNN                     & 0.122    & 0.033   & 0.202  & \textbf{0.760} & \textbf{0.346} & 1.224          \\
H\_Aug                       & 2D\_CNN\_SE                & 0.078    & 0.067   & 0.202  & \textbf{0.758} & \textbf{0.476} & 1.125          \\
I\_Aug                       & 2D\_CNN\_VSE                & 0.134    & 0.002   & 0.269  & \textbf{0.734} & \textbf{0.416} & \textbf{1.104} \\ \hline
\end{tabular}
\end{table*}

\begin{table*}[!h]
\caption{Fusion EER~(\%) results on SASV 2022 development and evaluation partitions}
\label{tab:fusion_results}
\renewcommand\arraystretch{1.2}
\centering
\begin{tabular}{cccccccc}
\hline
\multirow{2}{*}{Model Index}                                                                     & \multirow{2}{*}{Fusion Method}                                        & \multicolumn{3}{c}{DEV}     & \multicolumn{3}{c}{EVAL}                         \\ \cline{3-8} 
                                                                                                 &                                                                       & SASV-EER & SPF-EER & SV-EER & SASV-EER       & SPF-EER        & SV-EER         \\ \hline
\multirow{2}{*}{\begin{tabular}[c]{@{}c@{}}D\_Aug \& G\_Aug \& \\ H\_Aug \& I\_Aug\end{tabular}} & \begin{tabular}[c]{@{}c@{}} Linear Regression \end{tabular} & -        & -       & -      & 0.838          & 0.838          & 1.136          \\
                                                                                                 & Score Averaging                                                         & 0.067    & 0.067   & 0.135  & \textbf{0.559} & \textbf{0.354} & \textbf{0.857} \\ \hline
\end{tabular}
\end{table*}
To further improve the performance of the proposed systems, we add the ASVspoof~2019 development set to train the above models in Table~\ref{tab:results_train}. The results are illustrated in Table~\ref{tab:results_train_dev}. With the help of extra data in training, the SASV-EER of Extend(512)\_DNN model declines absolute 1.815\%. Moreover, the SASV-EERs of all the models are less than 0.998\% on the evaluation set except that of Extend(512)\_DNN model. At the same time, we find both the SPF-EER and SV-EER have further reduction. The best single model in this paper is the 2D\_CNN\_VSE, which is trained with  the training and development sets of ASVspoof 2019. The SASV-EER, SPF-EER and SV-EER of the best model are 0.734\%, 0.416\% and 1.104\% respectively. The results demonstrate the proposed embedding fusion methods, ensemble model structures and attention mechanisms have significant contributions to the backend ensemble of speaker verification and spoofing countermeasure systems.
 

Finally, we fuse the above models trained with the ASVspoof~2019 training and development set. We adopt two fusion methods, i.e., linear regression in Bosaris toolkit~\cite{brummer2013bosaris} and score averaging. The fusion results are shown in Table~\ref{tab:fusion_results}. After averaging the scores of the 1D\_CNN model and three 2D CNN-based models in Table~\ref{tab:results_train_dev}, the SASV-EER, SPF-EER and SV-EER reach 0.559\%, 0.354\% and 0.857\% respectively. We notice that the fusion of linear regression leads to worse results compared with those of score averaging. This is probably because the development set is used to train the models, which leads the models to overfit to the development set. Meanwhile, the experimental results on the development set from Table~\ref{tab:results_train} and Table~\ref{tab:results_train_dev} illustrate the models overfit to the development set when we add development set to train the models.


\section{Conclusions}
This paper develops a spoofing aware speaker verification system for the SASV challenge. In this challenge, we particularly focus on exploring the backend ensemble schemes from three aspects, which are embedding fusion methods, different ensemble frameworks and attention mechanism design respectively. Specifically, we propose circulant matrix transformation and stacking operation for embedding fusion. With the fused embeddings, we further introduce different backend ensemble frameworks based on CNN. Finally, we introduced parallel attention and SE attention in CNN-based frameworks to capture the global relationship between speaker embeddings and countermeasure embeddings. With the proposed method, the SASV-EER of our best single backend ensemble system reaches 0.734\% on the evaluation set. Together with the above contributions, the fusion on four well-trained models in this paper finally leads our submission to the fifth place in the SASV challenge.

\bibliographystyle{IEEEtran}

\bibliography{mybib}


\end{document}